\documentclass[aps,amsfonts,nofootinbib,twocolumn]{revtex4}
\usepackage{amsmath}

\begin{document}

\newcommand{\bb}{\begin{equation}}
\newcommand{\ee}{\end{equation}}
\newcommand{\eqb}{\begin{eqnarray}}
\newcommand{\eqf}{\end{eqnarray}}

\preprint{}
\title{Vortices in U(1) Noncommutative Gauge Fields }
\author{H.\ Falomir}
\email{falomir@fisica.unlp.edu.ar} \affiliation{IFLP -Departamento
de F\'{\i}sica, Facultad de Ciencias Exactas, Universidad Nacional
de la Plata, C.C. 67, (1900) La Plata, Argentina}
\author{J.\  Gamboa}
\email{jgamboa@lauca.usach.cl} \affiliation{Departamento
de F\'{\i}sica, Universidad de Santiago de Chile, Casilla 307,
Santiago 2, Chile}
\author{J.\  L\'opez-Sarri\'on}
\email{justo@dftuz.unizar.es} \affiliation{Departamento de
F\'{\i}sica, Universidad de Santiago de Chile, Casilla 307, Santiago
2, Chile}
\author{F.\ M\'endez}
\email{fmendez@lauca.usach.cl} \affiliation{Departamento de
F\'{\i}sica, Universidad de Santiago de Chile, Casilla 307, Santiago
2, Chile}
\author{A.\ J.\ da Silva}
\email{ajsilva@fma.if.usp.br } \affiliation{Instituto de
F\'{\i}sica, Universidade de S\~ao Paulo,
 CP  66316, 05315-970, S\~ao Paulo, SP, Brazil}

\begin{abstract}
 Charged  vortex solutions for  noncommutative Maxwell-Higgs  model in
 $3+1$ dimensions are found.  We show that the stability of these
   vortex   solutions  is   spoiled  out   for  some,   large  enough,
   noncommutativity parameter.  A non topological  charge, however, is
   induced by noncommutative effects.
\end{abstract}
\pacs{PACS numbers:}

\maketitle

%%%%%%%%%%%%%%%%%%%%%%%%%%%%%
%%%%%%%%%%%%%%%%%%%%%%%%%%%%%
\section{Introduction}
%%%%%%%%%%%%%%%%%%%%%%%%%%%%%
%%%%%%%%%%%%%%%%%%%%%%%%%%%%%

Lorentz  and CPT  symmetries are  cornerstone in  relativistic quantum
field theory which are cast on doubt in connection with recents experiments in
cosmic  physics  rays  \cite{GZK},  neutrino physics  \cite{LSND}  and
cosmological  measurements  \cite{WMAP}.  

In  order  to explain  these
experimental  results some authors  have suggested  conservative avenues, however 
explain  experiments  like  LNSD  and  some  other  astrophysical
observations apparently will require new and radical ideas.

By other hand vortex are localized, classical, finite energy solutions
that appears
in  some   quantum  field  theory   models  \cite{vinet,khare}.   Such
solutions  encode   the  collective   behavior  of  some   systems  at
nonperturbative  level.   Although   these  kind  of solutions
originally were found in  condensed matter physics \cite{abriko}, they
also  play an important  role in  particle physics  \cite{nielsen} and
cosmology,  in  the  language  of  cosmic  strings  \cite{kibble}  and
superstrings \cite{witt}.

A   particular  model  that   exhibits  this  solution   is  the
  Maxwell-Higgs  model in $3+1$  dimensions. It has also been shown
  \cite{khare} that properties  of vortices are modified if  we add to
  the theory  a Chern-Simons (CS)  term. For instance, they  acquire a
  finite quantized charge and angular momentum.

On the other hand, in previous papers \cite{us2,us3}, and from a
completely different  point   of  view,  we  have  reported   that a
particular  Lorentz invariance violation (LIV)  induces a CS term in
Abelian and non  Abelian Maxwell  theories. Our approach  to LIV
contains two scales, infrared  and ultraviolet,  and it is properly
defined in the field space rather than in spacetime.

Therefore, it is  natural to ask if our approach admits vortex-like
solutions and,  in this case,  which are their properties.

 In this paper we discuss some new features arising when an
infrared  cutoff is introduced  into the  Maxwell-Higgs model  in
$3+1$ dimensions, describing  the way  the  uncharged  vortex
solutions for the standard Maxwell-Higgs model are modified and
considering the stability of these solutions.

As discussed  in \cite{us1,us2,us3,loso} (see also \cite{jack}), an
explicit infrared cutoff appears in  quantum  field  theory  when
the canonical commutators of momenta are modified in analogy  with
the Landau problem in quantum mechanics.   The  presence  of  this
cutoff breaks  Lorentz symmetry and induces  a kind of dimensional
reduction which could provide a  new clue for many unsolved problems
in quantum field theory, cosmology and astrophysics.

Let  us  start reviewing  the  main  aspects  of quantum  theory
with noncommutative fields. Later, we will show that a theory with
noncommutative U(1)  fields coupled to a Higgs  scalar has indeed
vortex solutions.

In  Ref.   \cite{us1,us2,us3}, an  approach  to  a Lorentz
invariance violating  quantum   field  theory  has  been   proposed
inspired  in noncommutative geometry. The fields, instead of
satisfying the standard canonical commutators, obey \eqb
\left[\phi_i  ({\vec x}),  \phi_j({\vec y})\right]  &=&  i
\theta_{ij} \delta ({\vec x} -{\vec y}), \label{c1}
\\
\left[\pi_i  ({\vec x}),  \pi_j({\vec y})\right]  &=& i  {\cal B}_{ij}
\delta ({\vec x}-{\vec y}),
\label{c2}
\\ \left[\phi_i  ({\vec x}), \pi_j({\vec y})\right]  &=& i \delta_{ij}
\delta ({\vec x}-{\vec y}), \label{c3} \eqf where $i,j,...=1, 2,  3,
...$ are internal indices and $\theta_{ij}$   and   ${\cal B}_{ij}$
are   scales   with   dimensions   of $(\mbox{energy})^{-1}$ and
energy respectively.  For  small values of these scales, they
correspond to weak ultraviolet and infrared Lorentz invariance
violations respectively.

At this point, it is worth noting that this approach does not
correspond to the noncommutative  geometry in  the  usual sense,
where  one assumes  the commutator
    \[
    \left[x,y\right] \sim \theta.
    \]
Rather, while  the commutators (\ref{c1})  violates the
microcausality principle  imposing  an  ultraviolet  scale,
(\ref{c2}) affects  the physics in the infrared sector of the
quantum field theory \cite{kost}.

The noncommutative  scalar electrodynamics is defined  in full analogy
with  the electrodynamics, but  modifying appropriately  the canonical
commutators. 

The noncommutative scalar electrodynamics we will
consider in the present article is defined in full
analogy with  the electrodynamics, but  modifying
appropriately  the canonical commutators (For a model
based on the non-commutativity of coordinates and the
Moyal product for functions see \cite{mandal})

The Hamiltonian for this model becomes

\begin{eqnarray}
{\cal H} \,&=&\, \int d^3x\,\biggr\{\frac{1}{2}\left(E^2+B^2\right)
- A^0(\vec\nabla\cdot\vec E+\vec\theta\cdot\vec B)
\nonumber
\\
&+& (\Pi^*+ieA_0\phi)(\Pi-ieA_0\phi^*)-e^2A_0^2\phi^*\phi + e\Theta
\phi^* \phi A^0 \nonumber
\\
&+&\,  (\vec D \phi)^*\cdot(\vec D\phi) +
V(\phi^*\phi-\phi_0^*\phi_0)\biggr\}, \label{hamilncse}
\end{eqnarray}
where  $V(\phi^*\phi-\rho_0)$ is  a  scalar potential --
responsible for the symmetry breaking -- which can be written as
$$
V(\phi^*\phi-\phi_0^*\phi_0) = \frac{\lambda}{4}\left(\phi^*\phi -
\phi_0^*\phi_0\right)^2,
$$
with   $\lambda$  and  $\phi_0$ real and complex constants
respectively.

The modified commutators we adopt are
\begin{subequations}
\label{as}
\begin{eqnarray}
\left[E_i(\vec x),E_j(\vec y)\right] &=&
i\epsilon_{ijk}\theta^k\delta^3(\vec x-\vec y),
\label{a1}
\\
\label{a3}
\left[\Pi^*(\vec x),\Pi(\vec y)\right] &=& \Theta \delta^3(\vec
x-\vec y),
\end{eqnarray}
\end{subequations}
where $E_i$ is the electric field and $\Pi$ is the canonical
momentum associated to the charged  field $\phi$. Note that the
Gauss' law  is  modified  as in   (\ref{hamilncse}),  in  order  to
keep  gauge invariance with the modified commutators (\ref{a1})
and (\ref{a3}).

%%%%%%%%%%%%%%%%%%%%%%%%%%%%%%%%%%%%%%%%%%%%%%%%%%%%
%%%%%%%%%%%%%%%%%%%%%%%%%%%%%%%%%%%%%%%%%%%%%%%%%%%%
\section{Equation of motion and vortex solution}
%%%%%%%%%%%%%%%%%%%%%%%%%%%%%%%%%%%%%%%%%%%%%%%%%%%%
%%%%%%%%%%%%%%%%%%%%%%%%%%%%%%%%%%%%%%%%%%%%%%%%%%%%
In order to write  explicitly the equations of motion $(e.o.m.)$ for
this system, we will take  a coordinate system in which the third
axis $(z)$ is along the direction of the (space-like) vector
$\vec\theta$, that is, ${\vec \theta}=(0,0,\theta)$. Then, the
$e.o.m.$   for  the   Hamiltonian  (\ref{hamilncse})   with Poisson
brackets structure (\ref{as}), become
\begin{subequations}
\begin{eqnarray}
\partial_\mu F^{\mu\nu} +
\frac{\theta}{2}\epsilon^{3\nu\alpha\beta}F_{\alpha\beta} &=& J^\nu
- \delta^\nu_0\,e\Theta\phi^*\phi,
\\
D^\mu D _\mu\phi +\frac{\partial V}{\partial \rho} \phi+ i\Theta
D^0\phi &=& 0.
\end{eqnarray}
\end{subequations}
where, \eqb J^\mu &:=&  -ie\left[\phi^*(D^\mu\phi) -
(D^\mu\phi)^*\phi\right] \nonumber
\\
&=& -ie\left[ \phi^*\partial^\mu\phi-\phi\partial^\mu\phi^*\right]
-2e^2A^\mu\phi^*\phi,
\eqf
and
\[
D_\mu\phi= \partial_\mu\phi - ieA_\mu\phi,\quad (D_\mu\phi)^*=
\partial_\mu\phi^* + ieA_\mu\phi^*,
\]
with  $\rho:=\phi^*\phi$.

Using the Coulomb gauge condition, $\vec\nabla\cdot\vec A =
\partial_i A_i = 0,$ and considering static and $z$ coordinate independent
configurations, one obtains the equations
\begin{subequations}
\begin{eqnarray}
-\nabla^2 A^0  + \theta B_\theta = -e ( 2eA^0 &-& \Theta ) \phi^*\phi,
\\
-\nabla^2 A^i - \theta \epsilon^{ij}F_{0j} = J^i,\,\,\,\,i,j&=&1,2,
\\
-(\vec D)^2\phi +\frac{\partial V}{\partial \rho} \phi
- eA^0 (eA^0  -\Theta )\phi &=& 0,
\\
-\nabla^2A^3=-2e^2A^3\phi^*\phi,&&
\end{eqnarray}
\end{subequations}
which can be simplified by calling
\begin{subequations}
\begin{eqnarray}
\tilde A^0 &=& A^0 -\frac{\Theta}{2e}\\
\tilde{\phi}_0  &=& e^{i\zeta} \sqrt{\rho_0  - \frac{\Theta^2}
  {2\lambda}}\\
\tilde{\phi}^*_0 &=&
e^{-i\zeta}\sqrt{\rho_0-\frac{\Theta^2}{2\lambda}}
\end{eqnarray}
\end{subequations}
and redefining the scalar potential as
\begin{equation}
\tilde V(\rho) = V(\rho-\rho_0) + \frac{\Theta^2}{4}\rho.
\label{potencialmodificado}
\end{equation}

In such a way, the   set  of   equations   for  the   static
noncommutative scalar electrodynamics turns out to be,
\begin{subequations}
\label{eom}
\begin{eqnarray}
\label{ec-1}
-\nabla^2 \tilde{A}_0  &-& \frac{\theta}{2}\, \epsilon_{i j} F_{i j}
= -2e^2\tilde{A}_0 \phi^*\phi,
\\
\label{ec-2}
-\nabla^2 A_i &+& \theta \epsilon_{ij}\,\partial_{j} \tilde{A}_0 =
J_i,
\\
\label{ec-3}
-(\vec D)^2\phi + \frac{\lambda}{2}(\phi^*\phi
&-&\tilde{\phi}^*_0\tilde{\phi}_0)\phi - (e\tilde{A}_0)^2 \phi = 0,
\end{eqnarray}
\end{subequations}
which are just the equations for the static Maxwell-Chern
Simons-Higgs model \cite{schapo}.

But, contrarily to the commutative electrodynamics case, the linear
charge density operator is here shifted  according to
\begin{equation}
Q_\theta = \int d^2x\, J^0(x) = \tilde Q_\theta - e\Theta \int
d^2x\,\phi^*\phi
\label{shift}
\end{equation}
where,
$$\tilde Q_\theta \equiv \int d^2x\, \tilde{J}^0,~~~~~~~
~~~~~~~~~~~~~~~~~~~~~~~~~~~~~~~~~~~~~~
$$
$$
 = \int
d^2x\,\left\{-ie(\phi^*\partial^0\phi -\phi\partial^0\phi^*) -
2e^2\tilde{A}^0\right\}.
$$

 In order to analyze the asymptotic behavior of the fields we
choose cylindrical coordinates  $(r,\varphi,z)$ and follow
ref.
 \cite{khare}.  Thus, for $r\to\infty$ we have
\begin{subequations}
\begin{eqnarray}
\phi  &\rightarrow&   \tilde{\phi}_0  e^{i\alpha(\varphi)}+ {\cal
  O}(e^{-\lambda r})
\label{asympt} \\
A_i&\rightarrow& \frac{1}{e}\partial_i\alpha(\varphi)+  {\cal O} (
e^{-r/r_0})
\label{asympt1}\\
\tilde{A}_0&\rightarrow& \frac{1}{e}\partial_0\alpha(\varphi)+ {\cal
O} ( e^{-r/r_0}) = 0+ {\cal O} ( e^{-r/r_0}), \nonumber
\\
\label{asympt2}
\end{eqnarray}
\end{subequations}
with $r_0$  a constant with  apropriate dimensions. What  is
important here is the exponential behavior of  the corrections in
the limit we are considering.

On the other  hand, the requirement of $\phi $ to be single-valued
implies that $\alpha(2\pi)=  2 \pi n$, with $n$ an integer. A
suitable (smooth) gauge transformation  allows to choose
$\alpha=n\varphi$. Then, we conclude that the  magnetic flux is
quantized,
\begin{eqnarray}
\Phi_\theta = \int  d^2x\, B_\theta = \oint_{r\to\infty} d\vec
l\cdot \vec A = \nonumber \\
= \frac{1}{e}\left[\alpha(2\pi) - \alpha(0)\right] =
\frac{2\pi}{e}n\,. \label{flux}
\end{eqnarray}
But, as we will see later, the charge density is not proportional to
the flux as in the commutative case.

To solve the equations of motion, we  adopt  the usual Ansatz
\cite{nielsen}
\begin{subequations}
\label{ansatz}
\begin{eqnarray}
A_\varphi&=&
\frac{\sqrt{2\tilde\phi_0^*\tilde\phi_0}}{u}\,(n-g(u)),\\
A_0&=&\sqrt{2\tilde\phi_0^*\tilde\phi_0}\,h(u),\\
A_3&=& A_r = 0,\quad \phi = \sqrt{\tilde\phi_0^*\tilde\phi_0}\,
f(u)\, e^{in\varphi},
\end{eqnarray}
\end{subequations}
where $u=r\sqrt{2e^2\tilde\phi^*_0\tilde\phi_0}$.

Equations of motion (\ref{eom}), now reduces to
\begin{subequations}
\label{eom2}
\begin{eqnarray}
&&h^{\prime\prime}    +    \frac{1}{u}h^\prime    -    f^2\,    h =
\frac{\tilde\theta
  }{u}\, g^\prime \label{eqn16a}
  \\
&&g^{\prime\prime} - \frac{1}{u} g^\prime - f^2 g
  ={\tilde\theta }\, u\, h^\prime \label{eqn16b}
  \\
&&f^{\prime\prime} + \frac{1}{u}f^\prime -
  \frac{g^2}{u^2}\,f-\frac{\lambda}{4 e^2}
  (f^2-1) f=
-h^2 f\,, \nonumber
\\ \label{eqn16c}
\end{eqnarray}
\end{subequations}
where the primes stand for derivatives with respect to $u$ and
$\tilde\theta={\theta}\big/{\sqrt{2e^2\tilde\phi^*_0\tilde\phi_0}}$.

Equations (\ref{eqn16a}-\ref{eqn16c})  coincide    with   those
found in Ref.\ \cite{khare} in the context  of charged vortex
solutions. The energy per unit of length for this solution,
given by
\begin{eqnarray}
&&{\cal E}_n = 2\pi \tilde\phi_0^*\tilde\phi_0 \int_0^\infty
u\,du\,\left\{\frac{1}{
u^2}\left(\frac{dg}{du}\right)^2+\frac{1}{2 }\left(\frac{
 d f}{du}\right)^2 +\right.
\nonumber
\\
&+& \left(\frac{dh}{du}\right)^2+ \left.\frac{1}{2}
\left[\left(\frac{g}{u}\right)^2 +  h^2\right] f^2 + \frac{\lambda}{16
  e^2}\left(f^2(u) -1\right)^2\right\},
\nonumber
\\
&&
\label{linenergy}
\end{eqnarray}
turn out to be finite, up to an additive constant, for a suitable
class of boundary conditions.

 Remarkably, the  modified  potential
(\ref{potencialmodificado}) allows a direct  analysis  of the
conditions under  which  vortex  solutions exists  in the presence
of this kind of  noncommutativity. Indeed, we can rewrite
(\ref{potencialmodificado}) as
\begin{equation}
\tilde   V(\rho) = \frac{\lambda}{4}\left[\rho      -
\left(\rho_0-\frac{\Theta^2}{2\lambda}\right) \right]^2 +
\frac{\Theta^2}{4}\left(\rho_0 - \frac{\Theta^2}{4\lambda}\right)
\end{equation}
where the last term is  a physically irrelevant  constant and the
first one can be written as $V(\rho-\tilde\rho_0)$. Clearly, if
$\tilde\rho_0:= \rho_0-\frac{\Theta^2}{2\lambda} \leq  0$ then the
system does  not present the  symmetry breaking phenomenon and it is
not possible to find vortex-like solutions. Furthermore, for
$\tilde\rho_0=0$ one  can see that the energy  is zero.

Hence, one can  expect  that  in  order  to  find  finite  linear
energy density vortex-like solutions, the   condition
$\phi_0^*\phi_0     > \frac{\Theta^2}{2\lambda}$ should be
satisfied.

Concerning   the  linear charge density,   given   in
(\ref{shift}), we get from (\ref{ansatz})
\begin{equation}
\label{char}
Q_\theta    =   \frac{2\pi\theta}{e}n    -   \frac{\pi\,\Theta}{e}\int
du\,u\,f^2(u).
\end{equation}
The  last term  is independent  of $\tilde\phi_0$  and, since we
know from \cite{khare} that, for large $u$,
\begin{equation}
f(u)    \rightarrow    1    -    e^{-au},~~~~\mbox{with}~~~~a =
\frac{\sqrt{(\theta/2)^2-e^2\tilde\rho_0} - (\theta/2)} {\sqrt{2e^2
    \tilde\rho_0}},
\end{equation}
we can argue that the divergent part of  (\ref{char}) is a
contribution from  the  non trivial  vacuum with no physical
consequences. Indeed, the difference $\Delta_{nm}Q :=Q_n-Q_m$ is
finite (notice that the function  $f$  does not depend on
$n$) for any $n$ and $m$, including $n$ or $m$ equal to zero.

%%%%%%%%%%%%%%%%%%%%%%%%%%%%%%%%%%%%%%%%
%%%%%%%%%%%%%%%%%%%%%%%%%%%%%%%%%%%%%%%%
\section{Final remarks and conclusion}
%%%%%%%%%%%%%%%%%%%%%%%%%%%%%%%%%%%%%%%%%
%%%%%%%%%%%%%%%%%%%%%%%%%%%%%%%%%%%%%%%%%

Let us   summarize and  comment our results.  We have found that
there  exist   charged   vortex   solutions   in noncommutative
scalar   electrodynamics which  have similar properties to  those
found in   \cite{khare} for the commutative case. But in the present
case these solutions  depend on the noncommutativity parameters that
deform the scalar and vectorial momenta commutators.

As remarkable  differences with respect to the commutative case we
can point out that the charge of these vortices has not a
topological character and that, even for a non zero vacuum value of
the Higgs fields, there are no vortex solutions except  that $|
\phi_0 |^2$ be larger than a bound proportional to $\Theta^2$, where
$\Theta$ is the noncommutativity parameter of the Higgs field.

The point of  view advocated in this paper  is intimating related with
the Kostelecky {\it et al.} approach \cite{kost}, however a more precise
connection and, {\bf in particular,} the possible connection with the vortex
solutions found here is under research.

\section {Acknowledgements}

This work was supported from FONDECYT grants 1050114 (J.G), 1060079 (F.M) and 3060002 (J.L-S). The work of A.J.S was supported in part by FAPESP (Fundac\~ao de
Amparo a Pesquisa do Estado de S\~ao Paulo) and CNPq (Conselho
Nacional de Desenvolvimento Cient\'{\i}fico e Tecnol\'ogico). H.F acknowledge support from CONICET (PIP 6160) and UNLP (Proy.\
11/X381), Argentina.


\begin{thebibliography}{}
\bibitem{GZK}K.   Shinozaki  et   al.   (AGASA  Collaboration),   {\it
  Nucl.  Phys.  Proc.  Suppl.}{\bf  136},  18 (2004);  M.   Takeda  et
  al.  in {\it Hamburg 2001, Cosmic ray}, p. 341
(ICRC 2001).



\bibitem{LSND}C. Athanassopoulos et al. (LSND Collaboration), 
{\it Phys. Rev. Lett. } {\bf 81}, 1774 (1998).

\bibitem{WMAP} M. Bridges et al, astro/ph0607404.

\bibitem{vinet}  For  a review  on  vortex  solutions  see {\it  e.g},
  E. D'Hoker  and L. Vinet,  {\it Ann. of  Phys} (N.Y) {\bf  162}, 413
  (1985)

\bibitem{khare} A. Khare, {\it Proc.  Indian Natl. Sci. Acad}. {\bf
  A61}, 161, (1995).

\bibitem{abriko} A. A. Abrikosov, {\it Sov. Phys.} JEPT {\bf 5}, 1174 (1957).

\bibitem{nielsen} H. B. Nielsen and P. Olesen, {\it Nucl. Phys.} {\bf B61}, 45 (1973).

\bibitem{kibble} T. W. B. Kibble, {\it Phys. Rep.} {\bf 67}, 183 (1980).

\bibitem{witt} E.~Witten,
  %``Cosmic Superstrings,''
  Phys.\ Lett.\ B {\bf 153} (1985) 243.
  J.~Polchinski,
  %``Introduction to cosmic F- and D-strings,''
  arXiv:hep-th/0412244.
 J.~D.~Edelstein, C.~N\'u\~nez and F.~Schaposnik,
  %``Supersymmetry and Bogomolny equations in the Abelian Higgs model,''
  Phys.\ Lett.\ B {\bf 329} (1994) 39.
  J.~D.~Edelstein, C.~N\'u\~nez and F.~A.~Schaposnik,
  %``Supergravity and a Bogomolny bound in three-dimensions,''
  Nucl.\ Phys.\ B {\bf 458} (1996) 165.
  J.~D.~Edelstein, C.~N\'u\~nez and F.~A.~Schaposnik,
  %``Bogomol'nyi Bounds and Killing Spinors in d=3 Supergravity,''
  Phys.\ Lett.\ B {\bf 375} (1996) 163.
  M.~Shifman and A.~Yung,
  %``Localization of non-Abelian gauge fields on domain walls at weak  coupling
  %(D-brane prototypes II),''
  Phys.\ Rev.\ D {\bf 70} (2004) 025013.
  G.~Dvali, R.~Kallosh and A.~Van Proeyen,
  %``D-term strings,''
  JHEP {\bf 0401} (2004) 035.
  P.~Binetruy, G.~Dvali, R.~Kallosh and A.~Van Proeyen,
  %``Fayet-Iliopoulos terms in supergravity and cosmology,''
  Class.\ Quant.\ Grav.\  {\bf 21} (2004) 3137.
  J.~Lopez-Sarrion, E.~Moreno, F.~A.~Schaposnik and D.~Slobinsky, Phys.\ Rev. {\bf D73} (2006) 125007.

\bibitem{us2} J. Gamboa and J. Lopez-Sarrion, {\it Phys. Rev.} {\bf
D 71}, 067702 (2005).

\bibitem{us3}A. Das, J. Gamboa,  J. Lopez-Sarrion and F.A. Schaposnik,
  {\it Phys. Rev.{\bf D}72}, 107702 (2005).


\bibitem{loso} H. Falomir, J.  Gamboa, J. Lopez-Sarrion, F. Mendez and
  A. J. da Silva, {\it Phys. Lett.} {\bf B632}, 740 (2006).

\bibitem{us1}  J. M.  Carmona,  J.  L. Cort\'{e}s,  J.  Gamboa and  F.
  Mendez. JHEP {\bf  0303}, 058 (2003); {\it ibid},  {\it Phys. Lett.}
  {\bf B565} 222 (2003).


\bibitem{jack} S. Carroll, G. Field and R. Jackiw, {\it it Phys. Rev.} {\bf D41}, 1231 (1990);
A. A. Andrianov, P. Giacconi and R. Soldati, {\bf JHEP}, 0202
(2002).


\bibitem{us4}J. Gamboa, J. Lopez-Sarrion and A. P. Polychronakos,
{\it Phys. Lett. {\bf B}634}, 471-473 (2006).


\bibitem{kost} D. Colladay and V.A. Kosteleck\'y,  Phys. Lett. {\bf
B511}, 209 (2001); V.A. Kosteleck\'y, R. Lehnert, Phys. Rev. D {\bf
63}, 065008 (2001); R. Bluhm and  V.A. Kosteleck\'y, Phys. Rev.
Lett. {\bf 84}, 1381 (2000); V.A. Kosteleck\'y and  C. D. Lane,
Phys. Rev. D {\bf 60}, 116010 ( 1999); R. Jackiw and V.A.
Kosteleck\'y, Phys. Rev. Lett. {\bf 82}, 3572 (1999); D. Colladay,
V.A. Kosteleck\'y, Phys. Rev. D {\bf 58}, 116002 (1998); V. A.
Kosteleck\'y,  and  R. Potting, {\it Phys. Lett.} {\bf B381},89
(1996); R. Potting and R. Lehnert, {\it Phys. Rev. Lett.} {\bf 93},
110402 (2004); O. Bertolami, D. Colladay, V. A.  Kosteleck\'y and R.
Potting, {\it Phys. Lett.} {\bf B395}, 178 (1997).

\bibitem{mandal} D. P. Jatkar, G. Mandal, S. R. Wadia. CERN-TH-2000-193, MRI-P-000701, {\it JHEP} {\bf 0009}, 018 (2000). 

\bibitem{khare1}S. A.   Paul and  A.  Khare, {\it  Phys. Lett.  } {\bf
  B174}, 420 (1986), Erratum-ibid. {\bf 177B}, 453 (1986).

\bibitem{schapo} G. Lozano, M. V. Manias and F. A. Schaposnik {\it
Phys. Rev. D} {\bf 38}, 601 (1988); Zheng-Min XI {\it Phys. Lett.}
{\bf 217}, 113 (1988); Choonkyu Lee, Kimyeong Lee and Hyunsoo Min
{\it Phys. Lett. } {\bf 252}, 79 (1990).

\end{thebibliography}
\end{document}